# An Electrical Spinning Particle in Einstein's Unified Field Theory

S.N. Pandey[1], B.K. Sinha[2], Raj Kumar.[2]

1. Department of Mathematics, M.N.N.I.T, Allahabad (U.P.) India.

2. Department of Mathematics, U.N.S.I.E.T, V.B.S.P.U. Jaunpur (U.P.) India.

**ABSTRACT**

Previous work on exact solutions has been shown that sources need to be appended to the field equations of Einstein's unified field theory in order to achieve physically meaningful results, such sources can be included in a variational formulation by Borchsenius and Moffat. The resulting field equations and conservation identities related to the theory that can be used to derive the equations of structure and motion of a pole-dipole particle according to an explicitly covariant approach by Dixon[6]. In this present paper it is shown that, under certain conditions for the energy tensor of the spinning particle, the equations of structure and motion in an electromagnetic field turn out to be formally identical to those occurring in Einstein-Maxwell theory.

**Introduction :** In recent years it has been possible to find a class of exact solutions to the field equations of Einstein's unified field theory[7] that depends on three co-ordinates[1]. Particular solutions belonging to that class appear endowed with straight forward physical meaning as soon as sources are allowed for at the right hand side of the field equations. In this way the direct evidence is found, suggesting that the theory accounts for interactions not depending on distance between partons residing on a string, as well as for electromagnetic interactions[2]. Sources seem therefore an essential completion to the original field equations, as it occurs in general relativity, of which Einstein's unified field theory is a natural generalization. For such sources, a much desirable theory is lacking at present; they can be introduce phenomenologically in the form of a four current density $\tilde{j}^k$ and of a non-symmetric stress energy momentum tensor density $r_{\sim k\ell}$, if one adopts a variational formulation originated in the context of Bonner's theory[4] and subsequently extended to Einstein's unified field theory itself[8].

The evidence coming from the exact solutions intimates that it should be possible to derive from the field equations obtained in this way, from conservation identities, the equations of structure and motion of an electrical particle i.e. a particle that feels Lorentz force when moving in an electromagnetic field. Through the use of a non explicitly covariant method developed for general relativity[9], this possibility has been verified in the affirmative for a pole test particle, under certain conditions for the inner structure of the test body[3]. We know prove that the some conditions, when imposed to a pole – dipole a particle leads, through an explicitly covariant approach first introduced for general relativity[6], to equations of structure and motion that are formally identical to those occurring in Einstein-Maxwell theory.

**Field Equations and Conservative Identities :**

Let us introduce a non symmetric fundamental tensor in a four dimensional continuum $g_{k\ell} = g_{\underline{k\ell}} + g_{\underset{\vee}{k\ell}}$, a non-symmetric affinity $W^k{}_{\ell m} = W^k{}_{\underline{\ell m}} + W^k{}_{\underset{\vee}{\ell m}}$, as well as phenomenological sources, represented by a non symmetric tensor density $r_{\sim k\ell}$ and by and by a four current density $\underset{\sim}{j}{}^k$.

Let us define the four vector $W_k \equiv W^\ell{}_{\underset{\vee}{k\ell}}$, the scalar density $\sqrt{-|g_{k\ell}|}$, and the contravarient tensor $g^{k\ell}$ according to the equations

$$g^{km} g_{\ell m} = g^{mk} g_{m\ell} = \delta^k_\ell \qquad (1)$$

With these identities, we build the scalar density

$$\underset{\sim}{L} = \underset{\sim}{g}{}^{k\ell} \overline{R}_{k\ell}(W) - 8\pi \, \underset{\sim}{g}{}^{k\ell} \, r_{\sim k\ell} + \frac{8\pi}{3} W_k \, \underset{\sim}{j}{}^k \qquad (2)$$

where

$$\overline{R}_{k\ell}(W) = W^s{}_{k\ell,s} - \frac{1}{2}(W^s{}_{ks,\ell} + W^s{}_{\ell s,k}) - W^t{}_{s\ell} W^s{}_{kt} + W^s{}_{k\ell} W^t{}_{st} \quad (3)$$

is the symmetrical, contracted curvature tensor introduced by Borchsenius[4]. From the density $\underset{\sim}{L}$, with the usual variational procedure that consider $g^{k\ell}$ and $W^k{}_{\ell m}$ as independent variables and by introducing the new affinity

$$\Gamma^k{}_{\ell m} = W^k{}_{\ell m} + \frac{2}{3} \delta^k{}_\ell W_m \quad (4)$$

One eventually gets the field equations [8]

$$\underset{\sim}{g}^{qr}{}_{,p} + \underset{\sim}{g}^{sr} \Gamma^q{}_{sp} + \underset{\sim}{g}^{qs} \Gamma^r{}_{ps} - \underset{\sim}{g}^{qr} \Gamma^t{}_{\underline{pt}} + \frac{4\pi}{3} (\underset{\sim}{j}^r \delta^q{}_p - \underset{\sim}{j}^q \delta^r{}_p) = 0 \quad (5)$$

$$\underset{\sim,s}{\overset{ks}{g}}{}^{\vee} = 4\pi \underset{\sim}{j}^k \quad (6)$$

$$\overline{R}_{k\ell}(\Gamma) = 8\pi (r_{\underline{k\ell}} - \frac{1}{2} g_{\underline{k\ell}} r) \quad (7)$$

$$\overline{R}_{\underset{\vee}{k\ell}}(\Gamma) = 8\pi (r_{\underset{\vee}{k\ell}} - \frac{1}{2} g_{\underset{\vee}{k\ell}} r) - \frac{1}{3}(W_{k,\ell} - W_{\ell,k}), \quad (8)$$

where $\quad r = g^{k\ell} r_{k\ell} \quad (9)$

For a given $\underset{\sim}{j}^k$, under quiet general conditions, equation (5) univocally determines $\Gamma^k{}_{\ell m}$ in terms of fundamental tensor; Equation (6) expresses the non-homogeneous set of Maxwell's equations, while equation (8) just prescribes that $\overline{R}_{\underset{\vee}{k\ell}}(\Gamma) - 8\pi(r_{\underset{\vee}{k\ell}} - \frac{1}{2} g_{\underset{\vee}{k\ell}} r)$ is equal to a curl : $W_k$ remains infact unspecified, since it appears only in this equation. In a given co-ordinate frame, the field equations (5) to (8) can be uniquely solved for $g_{k\ell}$ under quiet general conditions, If $r_{\underset{\sim k\ell}{}}$ and $\underset{\sim}{j}^k$ are assigned; when both $g_{k\ell}$ and $r_{k\ell}$ are symmetric, these equations can be reduced to the equations of general relativity; When $r_{\underset{\sim}{k\ell}}$ and $j^k_{\sim}$ are both null, these can be reduced to the original equations of Einstein's unified field theory. The symmetrised Ricci tensor $\overline{R}_{k\ell}$ was adopted by Moffat[8] in order to preserve the transposition invariance of the original Einstein's equations even when sources were present, in the following sense: if $g_{k\ell}$ and $\Gamma^k_{\ell m}$ describe a solution to the field equations (5) to (8) for the given $r_{\underset{\sim}{k\ell}}$ and $j^k_{\sim}$, the transposed fields $\widetilde{g}_{k\ell} = g_{\ell k}$ and $\widetilde{\Gamma}^k_{\ell m} = \Gamma^k_{m\ell}$ give a solution to the same equations with sources $\widetilde{r}_{\underset{\sim}{k\ell}} = r_{\underset{\sim}{\ell k}}$ and $\widetilde{j}^k_{\sim} = -j^k_{\sim}$.

From the invariant integral

$$I = \int \left[ g^{k\ell}_{\underset{\sim}{}} \overline{R}_{k\ell}(W) + \frac{8\pi}{3} W_k j^k_{\sim} \right] d\Omega \tag{10}$$

By the well known method of infinitesimal co-ordinate transformation, and by using the field equations (5) to (8), one gets the weak identities

$$(g^{ks} r_{\underset{\sim}{k\ell}} + g^{sk} r_{\underset{\sim}{\ell k}})_{,s} + g^{pq}_{,\ell} r_{\underset{\sim}{pq}} + \frac{1}{3} j^k_{\sim}(W_{\ell,k} - W_{k,\ell}) = 0 \tag{11}$$

Let us call $\gamma_{k\ell}$ the symmetric part of $g_{k\ell}$ of the fundamental tensor, that the exact solutions[2] suggest to be the metric tensor of the theory and defines its inverse $\gamma^{k\ell}$ in the usual way. We further introduce the Christoffel symbol

$$\left\{ {k \atop \ell m} \right\} = \frac{1}{2} \gamma^{kp}(\gamma_{\ell p,m} + \gamma_{pm,\ell} - \gamma_{m\ell,p}) \tag{12}$$

And we pose

$$r^{pq}_{\sim} = g^{kq} g^{p\ell} r_{\underset{\sim}{k\ell}} \tag{13}$$

Then, weak identities (11) can be written as

$$\underset{\sim}{r}{}^{\ell s}{}_{|s} = \gamma^{\ell p} \underset{\underset{\vee}{g_{pq}}}{} \underset{\sim}{r}{}^{qs}_{\underset{\vee}{,s}} + \frac{1}{2} \gamma^{\ell p} g_{[qp,s]} \underset{\sim}{r}{}^{sq}_{\underset{\vee}{}} + \frac{1}{6} \gamma^{\ell p} \underset{\sim}{j}{}^{q}(W_{q,p} - W_{p,q}). \qquad (14)$$

where a vertical stroke is used to indicate that the covariant derivative with respect to Christoffel affinity (12) must be performed.

**An Electrical Pole-Dipole Particle :**

It should be clear, due to more form of the field equations (5) to (8), that from Einstein's unified field theory with sources one should not expect a mere account of gravitation and Maxwell's electromagnetism, with $g_{\underline{k\ell}}$ acting as metric field and $g_{k\ell}{}_{\vee}$ playing the role of electromagnetic tensor.

Relevant example of electromagnetic solution i.e. fields for which

$$g_{[k\ell,m]}{}_{\vee} = 0 \qquad (15)$$

have been found, it is also known that in general this equation can be satisfied up to the linear approximation by the theory[10], but of course, a truly unified description of nuclear and electromagnetic behaviour, intimated by the existence, side by side with electromagnetic examples of the string like solution mentioned in the introduction would be impossible if (15) were a field equation.

We want to as certain whether an electrical particle can be defined in the theory; therefore, we shall assume that equation (15) is indeed satisfied; this choice eliminates the second term of the right hand side of the equation (14). Of the two terms in the left, the first one describes a Lorents coupling between the electromagnetic field $g_{pq}{}_{\vee}$ and the conserved four-current density $\underset{\sim}{r}{}^{qs}_{\underset{\vee}{,s}}$, although this current density can not give rise to a net localized charge. The last term is intriguing, since it is the curl of a vector. $W_p$ appears there, Lorents coupled to the conserved four-current density $\underset{\sim}{j}{}^{q}$. We should remind however that the field equations

(5) to (8), have a peculiar feature at variance with what occurs in general relativity, a mere knowledge of the fundamental tensor is not enough to determine the energy tensor $r_{k\ell}$ univocally. The four current $j^i$ is fixed by equation (6), but equation (7) and (8) can not determine $r_{k\ell}$, due to the presence of the curl of the unknown vector $W_k$. This indeterminacy can be lifted if we pose, e.g.

$$\overline{R}_{k\ell}(\Gamma) = 8\pi r_{k\ell} \tag{16}$$

But, then, of course, equation (8) requires that

$$g_{k\ell} r = \frac{1}{12\pi}(W_{\ell,k} - W_{k,\ell}) \tag{17}$$

We consider equation (17) as a structure condition for an electrical particle in fact, when it holds the weak identities (14), with due regard to equation (15) takes the form

$$r^{\ell s}{}_{|s} = \gamma^{\ell p} g_{pq} J^q \tag{18}$$

where 

$$J^q = r^{qs}{}_{,s} + 2\pi r j^q \tag{19}$$

i.e. they show at right hand side is just a Lorent's term, with $g_{pq}$ acting as electromagnetic tensor, and $J^a$ as positive electric four - current density.

We remark that, in order to get equation (18) from (14) and (15), we only need that the extra injunction (17) holds wherever $j^q \neq 0$. Equation (18) is indeed a good result, but we shall ask ourselves whether and when the additional injunction (17) can be met with sources of the electrostatic solution[2] can indeed satisfy the injunction (17), but a general answer is obviously beyond reach. We note however that an easy way to fulfill equation (17) wherever $j^q \neq 0$ is at hand if r is constant there, since according to equation (15), we have already assumed that $g_{\underset{\vee}{ik}}$ is the curl of a vector.

Furthermore, if r is constant where $\underset{\sim}{j}^q \neq 0$, the conservation equation $(r \underset{\sim}{j}^q)_{,q} = 0$ is satisfied too, and the over all positive electric four - current density $\underset{\sim}{J}^a$ turns out to be a conserved quantity :

$$\underset{\sim}{J}^a{}_{,a} = 0 \qquad (20)$$

So sum up, we can obtain equation (18) and (20) by merely requiring that $g_{\underset{\vee}{k\ell}}$ is electromagnetic and that r is constant wherever $\underset{\sim}{j}^k$ is not vanishing. These two equations are formally identical to those occurring in Einstein–Maxwell theory; to them a covariant multipole formalism[6] can be applied without any change, provided that the world function, the bi-tensor of geodesic parallel displacement[11] and the covariant expansions[5] are all defined by using $\gamma_{k\ell}$ as metric and Christoffel symbol (12) as affinity. The resulting equations of structure and motion, in particular for the physically interesting case of an electrical spinning particle, turn out to be formally identical to those occurring in Einstein–Maxwell theory. It is remarkable, however, that in Einstein's unified field theory with sources the positive electric current density (19) and the active current density, defined by equation (6), are distinct entities. Quete note worthy is the role kept by the antisymmetric tensor $\underset{\sim}{r}^{\underset{\vee}{qs}}{}_{,s}$ can not be produce a net localized to the multi-pole electromagnetic structure of the particle, a physical meaning for the skew part of the energy tensor in Einstein's unifield field theory with sources begins to appear.

**Acknowledgement :**

The authors would like to thank Inter University Center for Astronomy and Astrophysics (IUCAA), Pune for providing facilities where this work was carried out. Finally this paper is devoted to Lord Shiva " Bhole Baba" .